\documentstyle[referee]{laa}

\begin{document}
\thesaurus{07  %A&A Section: Solar system
	   (07.09.1; % Interplanetary medium
%	    07.13.1;  % Meteoroids
%	    07.13.2   % Minor planets
	   )}

\title{Perihelion Concentration of Comets \\
       III. Physics and Astrophysics}

\author{ Jozef Kla\v{c}ka }
%\institute{Beniakova 20, 841 05 Bratislava,
%Slovak Republic}
\institute{Astronomical Institute,
Faculty for Mathematics and Physics,
Comenius University, \\
Mlynsk\'{a} dolina,
842 15 Bratislava,
Slovak Republic}
%E-mail: klacka@fmph.uniba.sk}
\date{}
\maketitle

\begin{abstract}
The problem of concentration of cometary perihelia on the sphere is
discussed from
the point of view of physics and astronomy. The important facts which may
be crucial in understanding the observed concentration of cometary perihelia
in the direction of solar apex are presented.
\end{abstract}

\section{Introduction}
Investigations of the last decades (see references in paper I of this series)
lead to the conclusion that perihelia of long-period comets exhibit
concentration in the direction of solar apex. The aim of this paper is to
discuss this result from the point of view of physics and astronomical
observations.

\section{Cometary Perihelia and Solar Apex -- Basic Properties}
If the observed concentration of cometary perihelia in the direction of solar
apex is of physical nature, then one should await that
the comets are of interstellar origin, i. e.,
the observed orbits
of the comets are characterized by osculating eccentricities mainly $e >$ 1.
No concentration in the direction of solar apex should be observed from the
physical point of view for the case of solar system origin of the bodies.

Important property pointing to the significance of the observed
concentration in a given direction consists in the simple fact that
the same direction (approximately) must be obtained for various subsets
of the fundamental input data (orbital elements).
These subsets may be of two types, in principle:
i) subsets defined by time intervals, e. g., the subset of the oldest known
bodies and the subset of the newest known bodies;
ii) subsets defined by geometrical conditions, e. g., the subsets are given
by various values of maximum value of perihelion distance.
If the observed concentration of cometary perihelia in the direction of solar
apex is of physical nature, then this direction of concentration must be
independent on a value of $q_{o}$, where $q_{o}$ is a given value defining
a subset of comets characterized by perihelion distances $q \le q_{o}$.

\section{Cometary Perihelia and Solar Apex -- Kinematics}
The observed concentration of cometary perihelia is in the direction of solar
apex. If the comets are of interstellar origin, then the physical contents
of the term ``solar apex'' is given by galactic astronomy and can be applied
on our case. We will do this application, now.

Let us consider the situation when the velocity vectors
of comets far away from the Sun, in the
interstellar space, are uniformly distributed with respect to the
reference frame rotating around the galactic center; the origin of the
frame corresponds to the place of the Sun. Any subset of the comets defined
by a small volume in the interstellar space near the Sun yields zero
final velocity vector (defined for the center of mass) with respect to the
reference frame, and, the motion of the Sun with respect to the subset of comets
is consistent with the term solar apex. Since
the observed concentration of cometary perihelia is in the direction of solar
apex, more comets must be situated behind the Sun than in front of the Sun
(the term `in front of' means that radii vectors of comets with respect to the Sun
%are situated in the direction of solar motion -- solar apex).
are of the same orientation as is the direction of solar motion --
orientation toward the solar apex).

Let us consider the situation when the radii vectors
of comets far away from the
Sun, in the interstellar space, are uniformly distributed with respect to the
Sun -- concentration of comets is independent on position with respect to the
Sun. Since the motion of the Sun with respect to the comets corresponds to
solar apex, the motion of the comets with respect to the Sun is in the direction
of antapex -- the perihelia of the comets are situated in the direction
of antapex. This is not consistent with observations --
the perihelia of the comets are situated in the direction
of apex. Thus, at least one of the assumptions -- interstellar origin of
comets, concentration of comets is independent on position with respect
to the Sun, solar apex determined from comets is consistent with solar
apex determined from various types of stars -- is not fulfilled.

\section{Cometary Perihelia and Solar Apex -- Astronomy}
Astronomical access to the result given by observational data
(observed concentration of cometary perihelia is in the direction of solar
apex) tries to consider observational selection effects which could influence
the observed concentration.
%Two important and well-known observational
%selection effects (manifested not only in cometary data, but also in data
%on asteroids) may play decisive role. The first effect corresponds to the
%fact that observations were made mainly from the northern hemisphere
%of the Earth. The second observational selection effect reflects the situation
The well-known observational
selection effect -- manifested not only in cometary data, but also in data
on asteroids -- may play decisive role.
The observational selection effect reflects the situation
that observations were made at favourite observational conditions --
at sites Europe (northern hemisphere of the Earth) prevails observations from
may to september.
%The combination of the both observational
%selection effects can play an important role in our understanding of the
%observed concentration of cometary perihelia is in the direction of solar apex.

\section{Conclusion}
We have presented physical and astronomical considerations which can play
an important role in our understanding of the
observed concentration of cometary perihelia in the direction of solar apex.
All these considerations must be taken into account when dealing with the data
and making final statements on the significance of the concentration
of cometary perihelia in the direction of solar apex.
One must try to find which of the discussed possibilities can be found
from the data. Application of the paper is presented in other parts
of this series of papers.

\acknowledgements
Special thanks to the firm ``Pr\'{\i}strojov\'{a} technika, spol. s r. o.''.
This work was also partially supported by Grant VEGA No. 1/4304/97.
\end{document}